\documentclass[a4paper]{article}

\usepackage{icrc2013,xspace,amssymb,url,breakurl}

\title{Characterization of the candidate site for the Cherenkov Telescope Array at the Observatorio del Teide}

\shorttitle{Characterization candidate site for CTA at the OT}

\authors{
I. Puerto-Gim\'enez$^{1}$, 
M. Gaug$^{2,3}$,
R. Barrena$^{1}$, 
J. Castro$^{1}$, 
M. Doro$^{2,3,4}$,
Ll. Font$^{2,3}$, 
M. Nievas Rosillo$^5$, 
J. Zamorano$^6$,
for the CTA Consortium.
}

\afiliations{
$^1$ Instituto de Astrof\'isica de Canarias, 38205 La Laguna, Tenerife, Spain   \\
$^2$ F{\'i}sica de les Radiacions, Departament de F{\'i}sica, Universitat Aut{\`o}noma de Barcelona, 08193 Bellaterra, Spain. \\
$^3$ CERES, Universitat Aut{\`o}noma de Barcelona-IEEC, 08193 Bellaterra, Spain. \\
$^4$ University and INFN Padova, Via Marzolo 8, 35131 Padova, Italy. \\
$^5$ Depto. F\'isica At\'omica, Universidad Complutense de Madrid, 28040 Madrid, Spain \\
$^6$ Depto. Astrof\'isica y Ciencias de la Atm\'osfera, Universidad Complutense, 28040 Madrid, Spain \\
}

\email{ipuerto@iac.es}

\abstract{The Spanish partners of the future Cherenkov Telescope Array
  (CTA) have selected a candidate site for the Northern installation
  of CTA, at 3 km from the Observatorio del Teide (OT), in the Canary
  Island of Tenerife. As the OT is a very well-characterized
  astronomical site. We focus here on differences
  between the publicly accessible measurements from the OT observatory
  and those obtained with instruments deployed at the candidate
  site. We find that the winds are generally softer at the 
  candidate site, and the level of background light comparable to the
  Observatorio del Roque de los Muchachos (ORM) at 
  La Palma in the B-band, while it is only slightly higher in the V-band.}


\begin{document}
\maketitle

\section{Introduction}
The Canaries have two islands with world-class skies for astronomy,
which can compete for hosting CTA-North. These islands host two
international observatories which together constitute the most
important optical, infrared and gamma-ray observatories in Europe: the
Observatorio del Teide (OT) at the island Tenerife, and the Observatorio del
Roque de los Muchachos (ORM) at the island La Palma. A specific
National Law protects the quality of the sky in these 
observatories~\cite{ley}. 
As La Palma does not fulfill the current CTA requirement of having
flat land of 1 km$^2$ above 1,500 meters, we have selected a site in
Tenerife, close to the OT.
This site is a plateau at an altitude of 2,260~m
a.s.l. with the geographic coordinates 28$^\circ$16'36''~N and
16$^\circ$32'08''~W. It has the great advantage of being located at
only 3~km from the OT, which has all kind of infrastructure for an
observatory available, and only 5~km away from the Iza\~na Atmospheric
Observatory (IZO), where the atmosphere is extensively characterized. 

The astronomical quality of an observatory is largely determined by
the transparency of the sky above it and the number of useful hours of
observation it offers each year (useful time). These two constraints
are intimately bound up with the site climate and geography.  
The excellent conditions at the region of the Canarian observatories
result from the Canary Islands geographical location,  
their orography, and the legal measures in place to protect sky quality. It's
worth mentioning that OT has been used and characterized as observatory for the
last 50 years, and that the characterization of the atmosphere at IZO covers
almost one century. This long-term data monitoring provides reliability to the
data presented here.

\begin{figure}[h!t]
\centering
\includegraphics[width=0.49\textwidth]{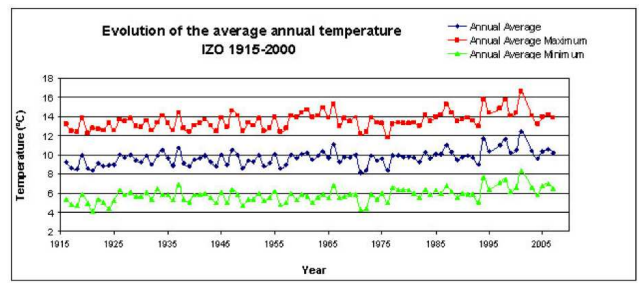} 
\vspace{-0.5cm}
\caption{\label{fig1}The evolution of the average annual temperature,
  average annual minima and maxima at
  IZO~\protect\cite{iarc}.
\vspace{-0.5cm}} 
\end{figure}

\vspace{-3mm}
\section{Weather Conditions}

%\subsection{Temperature Measurements}
{\bf Temperature measurements.}  
The average temperature at the OT is 9.8$^\circ$C, with 
minimum and maximum averages being 5.9$^\circ$C and 13.6$^\circ$C
respectively~\cite{temp}. These
agree with the temperatures measured at IZO,
yielding an average annual temperature around 10$^\circ$C, for the
period from 1915 to 2000, and average minima and maxima around
6$^\circ$C and 13$^\circ$C, respectively (Fig.~\ref{fig1}).\\  

\begin{figure}[h!t]
\centering
\includegraphics[width=0.48\textwidth]{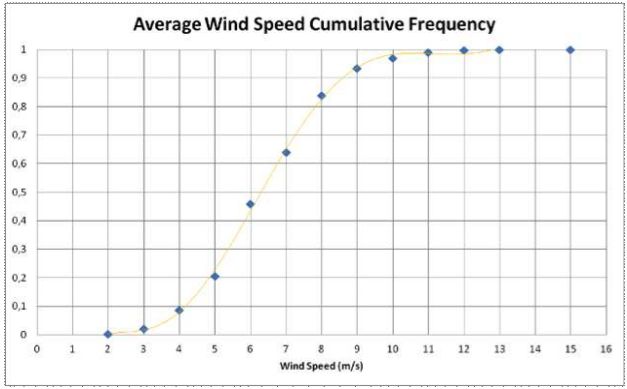} 
\vspace{-0.5cm}
\caption{\label{fig2} Cumulative monthly average wind speed
  frequency, as observed at IZO~\protect\cite{iarc2} from Jan 1933
  to Aug 2011.
\vspace{-0.5cm} }
\end{figure}

\noindent
{\bf Wind Measurements.} Due to the abrupt orography of the island,
wind speed and direction greatly vary from place to
place. 
%\subsection{Wind Measurements}
Wind speed data at IZO are available since 1933 and are publicly
accesible from the AEMET webpage~\cite{aemet}. The wind speed
measurements were taken every second or every 0.25 seconds (depending
on the anemometer which has changed over the years). The average wind
speed values have been averaged every 10 minutes. The IZO anemometer
was placed at different heights: from Jan 1933 to Feb 1984 at 12~m height,   
then until Feb 2000 at 16~m and then to present at 10~m height. 
Fig.~\ref{fig2} presents the monthly average wind speed cumulative 
frequency at IZO from 1933 to present: For over 90\% of the values, the
average wind speed is lower than 9~m/s (32.4~km/h). The average value
lies at 6.8~m/s (24.6~km/h). It is however expected that the wind
speeds are lower at the CTA candidate site (CTA-s), since IZO is located at
the top of the most exposed peak in the area, whereas the CTA-s is
more shielded.

\begin{figure}[h!t]
\centering
\includegraphics[width=0.48\textwidth]{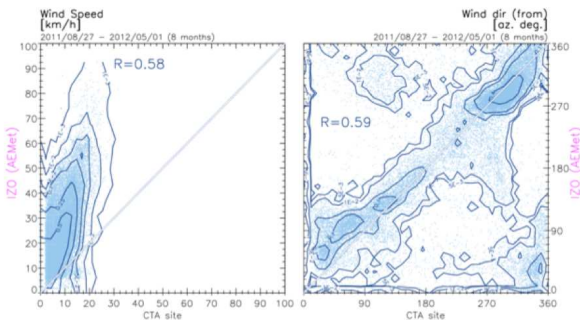} 
\vspace{-0.5cm}
\caption{\label{fig3}Simulatenous wind speed (left) and wind direction
(right) measurements at IZO and CTA-s, taken between Aug 2011 and Apr 2012.}
\end{figure}

In Fig.~\ref{fig3}, we show a comparison of simultaneous wind
measurements at IZO and the CTA-s, from Aug 2011 to Apr 2012. We compare wind at
10~m height at IZO with wind at 2.5~m height at the CTA-site
(Fig.~\ref{fig3}-left) and find that the wind speed at the
CTA-s is always about a factor 3 lower than at IZO. In
Fig.~\ref{fig3}-right we compare the simultaneous wind direction, and
find again that there is a quite strong global correlation between the
two sites.  

\begin{figure}[h!t]
\centering
\includegraphics[width=0.45\textwidth]{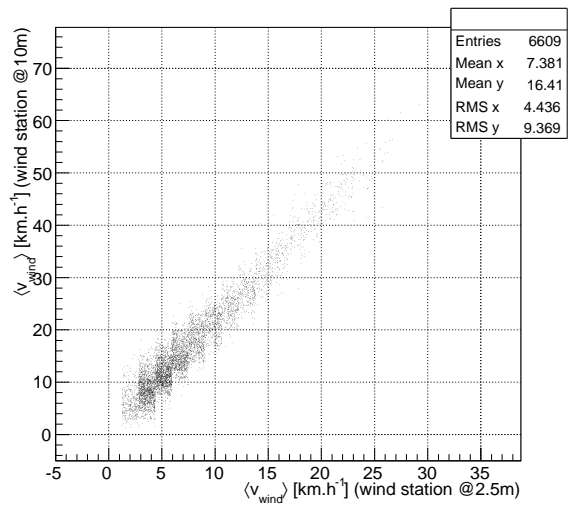} 
\vspace{-0.5cm}
\caption{\label{fig4} Comparison of wind speeds measurements at CTA-s
  taken with the anometer placed at 2.5~m and 10~m above the
  ground. Courtesy St. Vincent.} 
\end{figure}

To test whether the difference in the position of the anemometers can 
account for the large difference in wind speed observed at CTA-s
compared to IZO, we compare measurements taken both at CTA-s with the
sensor at 2.5~m and 10~m above the ground, from Dec 2012 to April
2013. Fig.~\ref{fig4} shows the results. We see a clear correlation,
and roughly an increase of a factor 2 in wind speed from 2.5~m to 10~m.  
Compared to Fig.~\ref{fig3} we still observe that the typical CTA-s
wind speed is about two thirds of that at IZO. \\

%\subsection{Humidity Measurements}
\begin{figure}[h!t]
\centering
\includegraphics[width=0.49\textwidth]{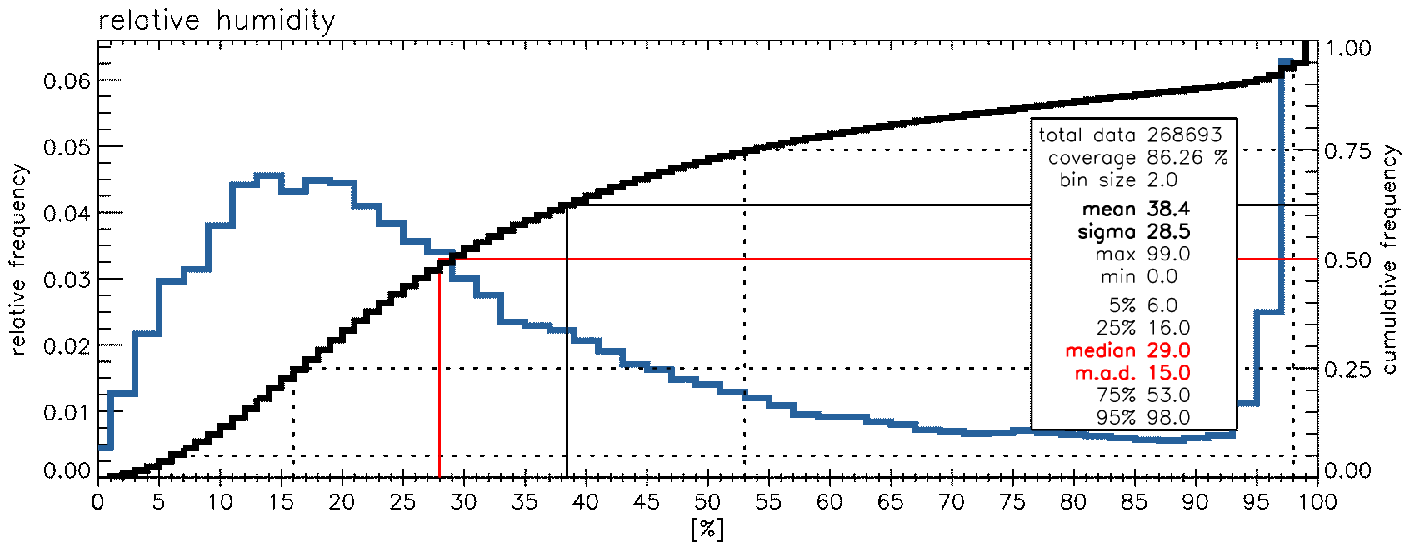} 
\includegraphics[width=0.49\textwidth]{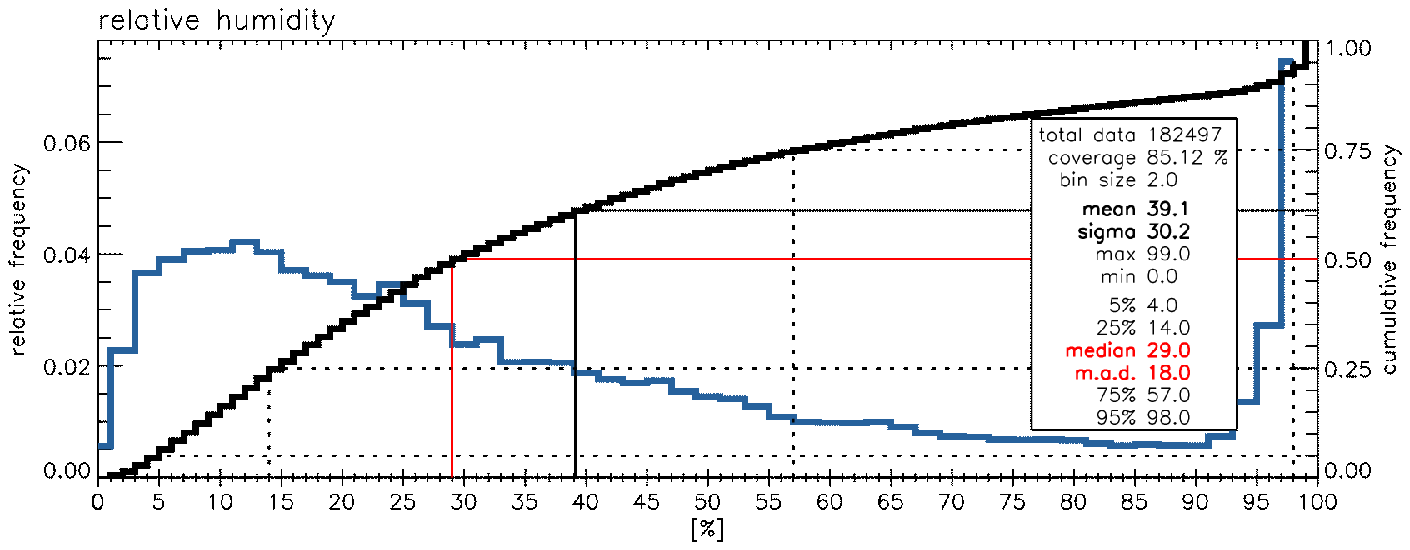} 
\vspace{-0.5cm}
\caption{\label{fig5} Frequency of
 daytime (top) and night-time  (bottom) 
humidity, as observed at 
%the OT
IZO
 during 
%one year. 
10 years.
%In black,
In blue,
  the normalized frequency while 
%in red
in black
 the cumulative distribution
  is shown. The sensor used for this campaign saturates at values
  around 
%90\%, 
99\%, 
which causes the abrupt change for the highest
  humidity values. These correspond to days with overspill of clouds,
  where the observatory is inside the clouds.
\vspace{-0.4cm}} 
\end{figure}

\noindent
{\bf Humidity Measurements.} 
The values presented here correspond to measurements taken at IZO and at the OT
%European Extremely Large Telescope (E-ELT) campaign carried out from May 2008 to May
%2009~\cite{elt}. 
The median, mean and percentiles (5\%, 25\%, 75\% and 95\%) are given 
%for day-time and night-time, separately 
in Fig.~\ref{fig5}. 
%The median is found at 27\% humidity, both for day
%and night, 
The median is found at 29\% humidity,
%and in 75\% of the night-time, the relative humidity is
%lower than 56\%. 
and in 75\% of the night-time, the relative humidity is
lower than 55\%. 
%The mean humidity value of 27\%, found for this one
The median humidity value of 29\%, found for this one
year of data, coincides with the value
obtained from a 5-year site characterization campaign for the
European Solar Telescope (EST), carried out from 2003 to 2008 at the
OT. However, the highest humidity values are less frequent in this
long database~\cite{sqg}.\\ 

%\subsection{Precipitation Measurements}
\begin{figure}[h!t]
\centering
\includegraphics[width=0.49\textwidth]{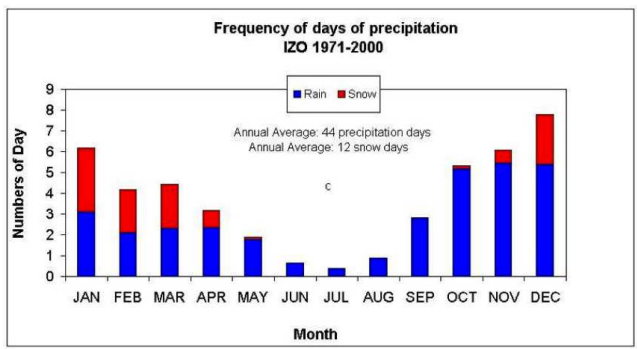} 
\vspace{-0.5cm}
\caption{\label{fig7} Frequency of precipitation-days per month,
  measured at IZO and averaged over 29 years~\protect\cite{iarc3}.
\vspace{-0.4cm}} 
\end{figure}

\noindent
{\bf Precipitation Measurements.}
The climate at the IZO region is extremely dry most of the year,
as it lies above the quasi-permanent temperature inversion layer, in
the dry free troposphere. Precipitation is mainly recorded in winter
time when Atlantic low areas pass over the Canary Islands. 
%In winter, the thermal inversion layer is weaker, and some days it may even
%disappear or is found above the IZO altitude.
%, resulting sometimes in fog and hoarfrost events. 
Fig.~\ref{fig7} presents the seasonal
frequency of days of appreciable precipitation (i.e. precipitation
$>$0.1~mm), observed at IZO for the period from 1971 to 2000.  
The annual average during this period is about 44 precipitation days.   
Out of these, about 11 days are of snow precipitation, mainly from Nov
to Apr, and, less frequently, from May to Oct. 
Considering only the days with precipitation $\gtrsim$10~mm, the
annual number of days goes down to 11 precipitation days per year. 
The AEMET public database also gives the number of days with hail per
month from 1920 to 2011. The annual average for this period amounts to
1.5 days of hail per year. \\ 

\vspace{-0.7cm}
\section{Atmospheric Quality}
The Cherenkov light that arrives to the telescopes from the particle
showers in the high atmosphere is highly dependent on the quality of
the atmosphere that it traverses. In fact different density profiles
lead to differences in Cherenkov light density of up to
60\%~\cite{bernlohr}, and the presence and position of clouds and 
aerosol layers also affect differently the atmosphere transmission~\cite{garrido}. \\ 

%\subsection{Cloud Coverage and Useful Nights}
\noindent
{\bf Cloud Coverage and Useful Nights.}
The atmosphere in the subtropical region of the Canary Islands is
characterized by its great stability throughout the year. This is due
to the combination of two processes of the atmospheric circulation at
large scale~\cite{palmen}. One of them is the descending branch of the
Hadley cell around 30$^\circ$~N and the other one the Trade or Alisios
Winds coming from the Azores high area that blow in the low
troposphere above an ocean which is relatively cold. As a result, a
temperature inversion layer appears around 1,300~m a.s.l. on average,
called the ``Alisio inversion'', which can be usually well identified
by the sea of stratocumulus on the Northern coasts of the islands.   
This layer separates two well-defined regimes: below it, there is the
moist marine boundary layer and above it, the dry free troposphere
(where the OT and the CTA-s are located). The Alisio inversion is a
quasi-permanent layer, being present 78\% of the time throughout the
year. Its altitude and thickness has a seasonal dependence, being
higher and thinner during the winter (when it is located between 1,350
and 1,850~m a.s.l., being only 350~m thick) and lower and thicker
during the summer between 750 and 1,400~m a.s.l., being about 550~m
thick)~\cite{torres}.  
Unfortunately, so far there are no specific and well-calibrated
instruments for ground measurements of cloud coverage at the OT or
IZO. We present here a compendium with the most relevant data found:  
sunshine measurements at IZO and useful time at the ORM observatory on
La Palma, 140~km away and also located at 2400~m a.s.l., i.e. also
above the inversion layer, and cloud coverage at the ORM calculated
with satellites. 

\begin{figure}[h!t]
\centering
\includegraphics[width=0.49\textwidth]{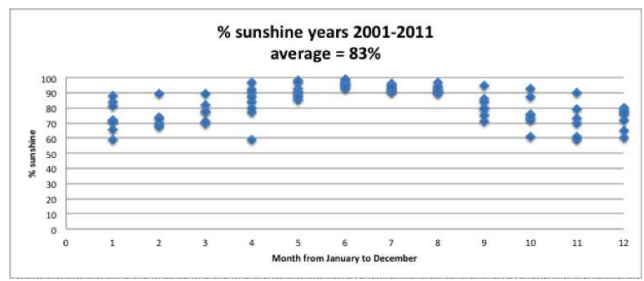} 
\vspace{-0.5cm}
\caption{\label{fig8} Monthly insolation, as observed at IZO~\protect\cite{iarc4}.
\vspace{-0.4cm}}
\end{figure}

The insolation, defined as the period of time during which the sun
shines, at IZO is very high, especially in summer, recording the
highest average annual insolation duration of Spain with
3448.5~hr/year. Fig.~\ref{fig8} shows the monthly insolation in
percentage taken with a heliograph at IZO from Jan 2001 to
Dec 2011. The average is 83\% of sunshine.
Detailed satellite studies have been carried out by the certified
consulting meteorologist Dr. Andr\'e Erasmus for the ORM on La Palma,  
showing that the photometric time at this observatory is 83.7\% (for
the 7-year period 1996-2002). Cross-calibration with ground
measurements using the Carlsberg Meridian Telescope (CMT) at the ORM
shows a good agreement of the data~\cite{erasmus}, with only 1.2\%
differences~\cite{casiana}. 
%Cavazzani et al.~\cite{cavazzani} use GOES
%satellite data for the time interval from 2007 to 2008 and find that
%the percentage of clear nights is 72.5\% at the ORM at La Palma.  

\begin{figure}[h!t]
\centering
\includegraphics[width=0.45\textwidth]{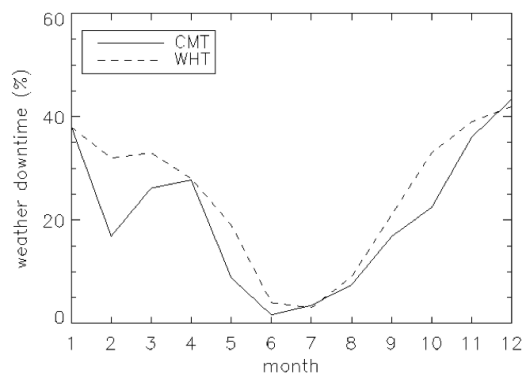} 
\vspace{-0.5cm}
\caption{\label{fig9} Relative monthly frequency of weather downtime
  at the CMT, averaged over 5 years, and compared with that of the WHT
  for an 18-year period~\protect\cite{garcia2}. 
\vspace{-0.0cm}} 
\end{figure}

%These numbers agree with 
The weather downtime calculated~\cite{garcia}
for the period 1999-2003 using the logs of the CMT telescope 
%This procedure 
obtained an average of 20.7\% weather downtime (defined as
the period when there were no recorded observations during a whole
night). The monthly average of weather downtime was also calculated
and compared with the one obtained at the William Herschel telescope (WHT) 
for the 18-year period from 1989 to 2006 and it is in good agreement
with the previous numbers, see Fig.~\ref{fig9}. 

Other in-situ measurements of useful time at the ORM yield the values:
78\%~\cite{murdin}, 75\%~\cite{ing} (11-year time interval) and
72.7\%-77.5\%~\cite{lombardi} (4-year period). More detailed
references and data can be found at the IAC Sky Quality Group
webpage~\cite{site}. At both observatories, the percentage of time
lost per year due to adverse atmospheric conditions remains between
16.3\% and 27.5\%, depending on the methodology for calculations and
the selected period of time. \\

%\subsection{In-site Particle Matter}
\noindent
{\bf In-site Particle Matter.}
Cuevas et al.~\cite{cuevas} concludes that in-situ TSP, PM10 data, and
Aerosol Optical Depth (AOD) observations performed at IZO demonstrate
that the site IZO-OT is characterized by extremely clean air and
pristine skies. IZO-OT is sometimes affected by dust-loaded African
air mass intrusions in the summer time (Jul-Sep), called
\emph{calima}. Since African air masses impact only in summer when
nights are short, the annual percentage of nocturnal observing
affected by calima is very low. In addition, most of the dust uses
to keep below the inversion layer and hardly affect the observatory.
Finally, the fact that under calima intrusions, middle clouds are normally
observed between 5 and 6~km altitude minimizes the negative impact of these
intrusions on astronomical observations performed at the OT. Overall, the
low impact of calima events can be seen in Fig.~\ref{fig8} where the weather
downtime percentage in summer months is lower than 10\% at the ORM.    \\ 

%\subsection{Night Sky Background}
\noindent
{\bf Night Sky Background (NSB).}
Here we present the results of a document about NSB at OT and ORM by
M. Nievas-Rosillo, supervised by R. Barrena-Delgado, from summer 2012,
to be published soon. The following table shows NSB measurements in
the V-band (mag/arcsec$^2$) for the OT taken with 3 different
instruments, an all-sky camera of type AstMon, an 80~cm telescope (the
IAC80) and the sky-quality-meter from Unihedron (SQM), similar to the
one used in the CTA-ATMOSCOPE, corrected for the light contamination
by the Milky Way:

\begin{center}
\begin{tabular}{lll}
Instrument  & All Data & From 2011 \\
\hline
AstMon (From 2012)  & 21,41 $\pm$ 0,15  & 21,41 $\pm$ 0,15  \\
IAC80 (From 2006)   & 21,24 $\pm$ 0,34  & 21,39 $\pm$ 0,38  \\
SQM (From 2006)     & 21,25 $\pm$ 0,23  & 21,21 $\pm$ 0,16  \\
\hline
\end{tabular}
\end{center}

The results of NSB measurements in the V- and B-bands
(mag/arcsec$^2$)~\cite{nievas} are compared between OT and ORM (both are 
telescope measurements) in the next table:

\begin{center}
\begin{tabular}{lll}
Instrument  & B & V \\
\hline
IAC-OT       & 22,34 $\pm$ 0,25   & 21,23 $\pm$ 0,13\\
ORM          & 22,70 $\pm$ 0,03   & 21,90 $\pm$ 0,03\\
Diff OT-ORM  & 0,36 $\pm$ 0,24    & 0,67 $\pm$ 0,13 \\
\hline
\end{tabular}
\end{center}

We can see that the typical night sky at OT is brighter than that of ORM,
which is a very dark astronomical observatory. It is interesting to note
that the difference in the B band, which is closer to the spectral
sensitivity of the photo-multipliers employed for the CTA telescopes is only
$0,36\pm 0,24$~mag/arcec$^2$. \\

\vspace{-0.7cm}
\section{Conclusions}
The site proposed by the Spanish partners of CTA at 3~km from the OT consitutes
an excelent candidate to host CTA-North. Actually, OT, together with ORM in La Palma, is
a very extensively and long-term characterized astronomical observatory. 
Such a long-term database makes the knowledge of the atmospheric conditions very
reliable and reduces the risk of a wrong characterization from short-term measurements. We have
summarized the main climate and atmospheric parameters obtained from public databases.
In order to complement the public databases, some measurements have been taken with instruments
deployed at the candidate site. Of special importance is the result we have obtained by comparing
wind speeds: the wind speed at the candidate site is about one thirds lower than that at IZO, according to the
fact that the site is more shielded than the IZO anemometer. Recent results on the NSB measurements
at the OT compared with ORM show that even though the night sky at OT is brighter than that of ORM,
the difference in the B band is very low and will hardly affect the CTA performance.   

\vspace{0.25cm}
{\bf Acknowledgements:\xspace} 
We gratefully acknowledge support from the agencies and organizations 
listed in this page:\\
\url{http://www.cta-observatory.org/?q=node/22}
\vspace{-0.25cm}

\bibliographystyle{plain}

\end{document}